\documentclass[final,5p,times,twocolumn]{elsarticle}
\usepackage{csquotes}
\usepackage{amsmath}
\usepackage{slashed}
\usepackage{bm}
\usepackage{caption}
\usepackage{booktabs,dcolumn}
\usepackage[colorlinks=true, linkcolor=blue, citecolor=blue, urlcolor=blue]{hyperref}
\usepackage{orcidlink}

\usepackage{amssymb}
\usepackage{lipsum}

\journal{Physics Letters B}

\begin{document}

\begin{frontmatter}



	\title{Neural-network excited states of $A=4$ nuclei and hypernuclei}

\author[FD]{Zi-Xiao Zhang~\orcidlink{0009-0003-4773-1505}}

\author[PK]{Yi-Long Yang~\orcidlink{0000-0002-5065-1309}}

\author[phys]{Xiao-Lu Qian~\orcidlink{0009-0009-2096-4847}}

\author[FD, IMP]{Wan-Bing He~\orcidlink{0000-0002-3854-4965}\corref{cor1}}
\ead{hewanbing@fudan.edu.cn}

\author[PK]{Peng-Wei Zhao~\orcidlink{0000-0001-8243-2381}}

\author[G]{Bing-Nan Lu~\orcidlink{0000-0001-7890-4948}}

\author[FD, IMP, ECNU]{Yu-Gang Ma~\orcidlink{0000-0002-0233-9900}\corref{cor1}}
\ead{mayugang@fudan.edu.cn}

\cortext[cor1]{Corresponding Author}
\address[FD]{Key Laboratory of Nuclear Physics and Ion-beam Application (MOE), Institute of Modern Physics, Fudan University, Shanghai 200433, China}
\address[PK]{State Key Laboratory of Nuclear Physics and Technology, School of Physics, Peking University, Beijing 100871, China}
\address[phys]{Department of Physics,
Fudan University, Shanghai 200433, China}
\address[IMP]{Shanghai Research Center for Theoretical Nuclear Physics, NSFC and Fudan University, Shanghai 200438, China}
\address[G]{Graduate School of China Academy of Engineering Physics, Beijing 100193, China}
\address[ECNU]{School of Physics, East China Normal University, 200062, Shanghai, China}

\begin{abstract}
We present the first variational Monte Carlo study of nuclear and hypernuclear excited states within the neural-network quantum states (NQS) framework.
We implement both the overlap penalty (OP) and natural excited state (NES) methods to compute low-lying excitation spectra. To address the spin contamination in hypernuclear calculations, we propose a quantum number targeting (QNT) technique for the OP method. Both the OP-QNT and NES methods can reproduce diagonal observables, such as energies and spatial structures, in excellent agreement with rigorous benchmarks. We further provide, to our knowledge, the first \textit{ab initio} calculation of the $M1$ transition strength for $^{4}_{\Lambda}\mathrm{H}$. The calculated transition strength is consistent with the weak-coupling limit, exhibiting a $\sim$1.3\% suppression.  This work demonstrates that NQS can be elevated from ground-state solvers to practical tools for nuclear and hypernuclear spectroscopy.

\end{abstract}




\end{frontmatter}




\section{Introduction}
\label{Section:Intro}

The $A=4$ nuclear and hypernuclear systems, such as $^4$He and $^4_\Lambda$H, are widely considered essential first benchmarks and testbeds for novel \textit{ab initio} few-body methods and underlying baryon-baryon interactions. Historically serving as a cornerstone of the nuclear
shell model, the $\alpha$ particle is the lightest doubly magic nucleus, featuring a highly stable and symmetric ground state with an excitation energy of roughly 20 MeV~\cite{PhysRevLett.132.062501}. In addition, the first excited state of  $^4$He shares the same quantum number, $0^+$, with its ground state.
Consequently, the spectrum of $^4$He cannot be extracted simply by leveraging the projection operators. 
Going beyond the first generation quark sector, the spectrum of $^4_\Lambda$H also introduces critical insights into hyperon-nucleon ($YN$) interactions. The energy splitting between its $0^+$ and $1^+$ states provides the most definitive constraints on the spin-dependent components of the $YN$ interaction \cite{RevModPhys.88.035004}. Furthermore, high-precision 
$\gamma$-ray \cite{BEDJIDIAN1976467} and decay-$\pi$ \cite{PhysRevLett.114.232501} spectroscopy of the $A=4$ hypernuclear isodoublet ($^4_\Lambda$H and $^4_\Lambda$He) have revealed a remarkably large, spin-dependent charge-symmetry-breaking (CSB) effect \cite{PhysRevLett.115.222501}.
Thus, accurate \textit{ab initio} calculations of the low-lying excited states in these 
$A=4$
systems remain both physically important and technically challenging.

Among the diverse applications of machine learning techniques in physics \cite{RevModPhys.91.045002, RevModPhys.94.031003, Lovato:2026erx, Giri:2025pkw, Wang2026,45g7-bmp6, PhysRevResearch.2.043202, 4ccs-66c6,PhysRevC.105.L031303,PhysRevC.111.034329,Wu:2023pzn,He:2023urp,He:2023zin, Liu2024}, the representation of wavefunctions by artificial neural-networks has achieved remarkable success in solving quantum many-body problems ~\cite{neuralsci}. In recent years, the neural-network quantum states (NQS) have been successfully adapted to investigate atoms and molecules~\cite{Ferminet,FermiSci,PauliNet, Laplacian, deepwf, Entwistle2023,Wheeler_2024,Symmetries}, ground states of atomic nuclei~\cite{v2007,Hidden,A6,Feynmannet,distill,cpl_42_5_051201, tlqz-nw28, ygkf-llyp, yang2025zemachradiinuclearstructure}, hypernuclei~\cite{ZHANG2026140285, wmxg-cnrg} and multiquark bound states~\cite{ckpr-s876}, ultracold Fermi gas~\cite{Kim2024}, homogeneous electron gas~\cite{PhysRevX.14.021030} and the crust of neutron stars~\cite{PhysRevResearch.5.033062, Fore2025}, with the parameters optimized by variational Monte Carlo (VMC) method. Within this VMC-NQS framework, provided that the ansatz is sufficiently expressive, the optimization process recovers the exact imaginary-time projection~\cite{PhysRevB.61.2599, Stokes2020}. Building upon the standard neural-network architecture, the integration of hidden nucleonic degrees of freedom~\cite{Hidden} and many-body backflow transformations~\cite{Feynmannet, ZHANG2026140285} further boosts the accuracy of VMC-NQS in \textit{ab initio} nuclear calculations, rivaling or even surpassing established methods such as hyperspherical harmonics (HH), auxiliary field diffusion Monte Carlo (AFDMC), and the stochastic variational method (SVM). Consequently, an intriguing question is whether VMC‑NQS can achieve similar success for the excitation spectra of atomic nuclei and hypernuclei.

Compared to ground-state calculations, computing low-lying excited states requires additional orthogonal constraints beyond the variational process alone. Within the VMC framework, imposing symmetry constraints generally relies on two fundamental strategies: modifying the target Hamiltonian (the variational loss function) and intrinsically reconstructing the ansatz. Corresponding to these two paradigms, two primary methodologies have recently gained prominence in maintaining orthogonality among states. One is the overlap penalty (OP) framework~\cite{Entwistle2023,Wheeler_2024,Symmetries}, in which the orthogonality is achieved by explicitly penalizing the overlap between states. The other one is the recently proposed natural excited state (NES) method~\cite{FermiSci}. This method  equates the problem of finding the lowest $K$ excited states to finding the ground state of a $K$-Fermion system. In this way,  the variational collapse is prevented naturally by the Pauli blocking among Fermions. Despite their successful applications in quantum chemistry, excited-state spectroscopy within the VMC-NQS framework has not yet been explored in nuclear and hypernuclear systems.

To bridge this long-standing gap, we present the first application of excited-state VMC-NQS methods to \textit{ab initio} nuclear and hypernuclear spectroscopy, by systematically comparing the OP and NES approaches for the low-lying spectra of $^{4}\mathrm{He}$ and $^{4}_{\Lambda}\mathrm{H}$. In all calculations, we employ the message-passing neural-network (MPNN) architecture for the NQS and the improved stochastic reconfiguration (SR) algorithm for minimization, as detailed in Ref.~\cite{ZHANG2026140285}. To accelerate the convergence, we adopt the Adam optimizer for pretraining. We further propose the quantum number targeting (QNT) technique to avoid the severe spin contamination in the calculations of $^{4}_{\Lambda}\mathrm{H}$. Equipped with these state-of-the-art excited-state methods, we report, to our knowledge,  the first \textit{ab initio} prediction of the reduced magnetic dipole transition probability $B(M1)$ of $^{4}_{\Lambda}\mathrm{H}$. Our study paves the way for the systematic application of NQS to the excited-state spectroscopy of nuclei and hypernuclei.

\section{Excited-state methods}

\subsection{Overlap penalty method}
In this section, we discuss the overlap penalty method for excited-state calculations. The core philosophy of this method is to modify the standard energy expectation by  introducing a penalty term proportional to the overlap between states. 
Guided by this idea, to find the lowest $K$ eigenstates of Hamiltonian $H$, we can generally define a $K$-state functional of a trial wavefunction ensemble \{$\Psi_i$\}~\cite{Wheeler_2024,Symmetries}
\begin{equation}
    L[\{\Psi_i\}] = \sum_i w_i E[\Psi_i] + \lambda \sum_{i<j} |S_{ij}|^2,
    \label{oploss}
\end{equation}
where the energy expectation value $E[\Psi_i]$ and the pairwise overlap $S_{ij}$ between states $i$ and $j$ are defined as
\begin{align}
    E[\Psi_i] = \frac{\langle\Psi_i|H|\Psi_i\rangle}{\langle\Psi_i|\Psi_i\rangle},\quad 
    S_{ij} = \frac{\langle\Psi_i|\Psi_j\rangle}{\sqrt{\langle\Psi_i|\Psi_i\rangle\langle\Psi_j|\Psi_j\rangle}}.
\end{align}
 Here, $w_i$ and $\lambda$ are positive hyperparameters.
In the OP method, the key point is how to choose these hyperparameters.   

Starting from a randomly initialized ensemble of trial wavefunctions, employing uniform weights ($w_1=w_2=\dots$) leaves the optimization landscape symmetric with respect to state permutations. Under such conditions, the optimizer lacks a clear gradient signal to distinguish which specific trial state should converge to the ground state and which to the higher-lying excited states, often resulting in severe numerical instabilities. To explicitly guide the optimization process, it is a standard practice to assign the weights in a strictly decreasing order ($w_1 > w_2 >\dots$), which prioritizes the minimization of states with lower energies.

Next, we discuss the choice of 
$\lambda$. Without the penalty term ($\lambda=0$), the optimization of 
$L$ forces all states to converge  to the exact ground state. 
As 
$\lambda$ is increased, the  collapse between distinct states becomes increasingly penalized. By analyzing the gradient and Hessian matrix of the loss function, it can be shown that there is a critical value $\lambda_c$~\cite{Wheeler_2024}, 
\begin{align}
	\lambda_c = \max_{j>i}\left[ (E_{j}-E_i) \frac{w_i w_j}{w_i - w_j} \right],
\end{align}
beyond which the orthogonality constraint is sufficiently enforced. Theoretically, for any \(\lambda > \lambda_c\), the exact minimization of the functional \(L[\{\Psi_i\}]\) strictly yields the lowest \(K\) true eigenstates. 
The Monte Carlo evaluation of the $S_{ij}$ in Eq. (\ref{oploss}) and the corresponding gradients with respect to the variational parameters are detailed in the Supplementary Material.

\subsection{Natural excited states method}
\begin{table*}[t]
    \centering
    \caption{\label{Tab:He4_results} TABLE 1. Comparison of the ground state ($0^+_1$) and the first excited state ($0^+_2$) of $^4$He computed using the OP and NES methods. The table presents the energy ($E$), point-proton radius ($r_p$), expectation values of the squared orbital ($\langle \mathbf{L}^2 \rangle$) and spin ($\langle \mathbf{S}^2 \rangle$) angular momenta for each state, along with the excitation energy ($\Delta E$). The calculations are performed using the LO-$\slashed{\pi}$EFT Hamiltonian with two-body force only. The energies given by HH method~\cite{PhysRevC.100.034004} are displayed for comparison. Note that the NES method intrinsically avoids the hyperparameter $\lambda$.}
    \begin{tabular}{l c c c c c c c c c c}
        \toprule
        & & \multicolumn{4}{c}{Ground state ($0^+_1$)} & \multicolumn{4}{c}{First excited state ($0^+_2$)} & \\
        \cmidrule(lr){3-6} \cmidrule(lr){7-10} 
        Method & $\lambda$ [MeV] & $E$ [MeV] & $r_p$ [fm] & $\langle \mathbf{L}^2 \rangle$ & $\langle \mathbf{S}^2 \rangle$ & $E$ [MeV] & $r_p$ [fm] & $\langle \mathbf{L}^2 \rangle$ & $\langle \mathbf{S}^2 \rangle$ & $\Delta E$ [MeV] \\
        \midrule
        OP  & 30 & -39.46(2) & 1.30(1) & 0.00(0) & 0.00(0) & -12.12(6)  & 2.91(2) & 0.00(0) & 0.02(0) & 27.34(6) \\
        OP  & 35 & -39.75(0) & 1.21(0) & 0.00(0) & 0.00(0) & -11.08(1)  & 3.00(1) & 0.00(0) & 0.02(0) & 28.67(1) \\
        OP  & 500 & -39.75(1) & 1.22(0) & 0.00(0) & 0.00(0) & -10.72(2)   & 2.96(1) & 0.23(0) & 0.24(0) & 29.03(2) \\
        NES & -- & -39.77(0) & 1.21(0) & 0.00(0) & 0.00(0) & -11.09(0)  & 3.18(1) & 0.00(0) & 0.02(0) & 28.66(0)\\
        HH~\cite{PhysRevC.100.034004} & -- & -39.843 & -- & 0.00 & 0.00 & -11.193 & -- & 0.00 & 0.00 & 28.650 \\
        \bottomrule
    \end{tabular}
\end{table*}
As an elegant alternative to the penalty-based approaches, the NES method circumvents the need for explicit overlap calculations and hyperparameter tuning~\cite{FermiSci}. The essence of NES is rooted in the Pauli exclusion principle.  To intrinsically prevent variational collapse, the NES method casts the $K$ states as $K$ fictitious Fermions in an augmented space. The total ansatz $\Psi$ is defined as
a Slater determinant of single-state ansatzes $\psi_i$
\begin{align}
	\Psi(\mathbf{x}^1, \dots, \mathbf{x}^K) \triangleq \det 
	\begin{bmatrix}
		\psi_1(\mathbf{x}^1) & \cdots & \psi_K(\mathbf{x}^1) \\
		\vdots & & \vdots \\
		\psi_1(\mathbf{x}^K) & \cdots & \psi_K(\mathbf{x}^K)
	\end{bmatrix},
\end{align}
where $\textbf{x}^i$ denotes the complete set of coordinates, spins and isospins of the $i$-th replica system. 
The antisymmetric property of the determinant ensures that the 
$K$ single-state ansatzes remain strictly linearly independent throughout the optimization.
 
The Hamiltonian $\mathcal{H}$ acting on the total ansatz is defined as the direct sum of $K$ single-state Hamiltonians, $\mathcal{H} = H_1\oplus...\oplus H_K$. 
In what follows, we adopt the following notation
\begin{align}
	\textbf{N} \triangleq  
	\begin{bmatrix}
		\langle\psi_1|\psi_1\rangle & \cdots & \langle\psi_1|\psi_K\rangle \\
		\vdots & & \vdots \\
		\langle \psi_K|\psi_1\rangle & \cdots & \langle\psi_K|\psi_K\rangle
	\end{bmatrix},
\end{align}
\begin{align}
	\textbf{O} \triangleq  
	\begin{bmatrix}
		\langle\psi_1|O|\psi_1\rangle & \cdots & \langle\psi_1|O|\psi_K\rangle \\
		\vdots & & \vdots \\
		\langle \psi_K|O|\psi_1\rangle & \cdots & \langle\psi_K|O|\psi_K\rangle
	\end{bmatrix}.
\end{align}
Therefore, we have the optimization objective in the NES method to be
\begin{align}
	\frac{\langle\Psi|\mathcal{H}|\Psi\rangle}{\langle\Psi|\Psi\rangle} = \text{Tr}\left[\mathbf{N}^{-1}\mathbf{H}\right].
	\label{NES objective}
\end{align}
We give an alternative derivation of this relation in the Supplementary Material, based on Jacobi's formula~\cite{vein1999determinants}, complementing the original proof using the matrix determinant lemma~\cite{FermiSci}. 

Although the NES method intrinsically ensures that the 
$K$ single-state ansatz are linearly independent, these individual states do not directly correspond to the exact eigenstates. Instead, $\{\psi_i, i=1,...,K\}$ merely span the same 
$K$-dimensional subspace as the lowest 
$K$ eigenstates $\{\phi_i\}$.
To recover the true energy spectrum $\mathbf{\Lambda}$, we need to diagonalize the resulting energy matrix 
\begin{align}
	\mathbf{N}^{-1}\mathbf{H} = \mathbf{U}\mathbf{\Lambda}\mathbf{U}^{-1}.
\end{align}
In the above, $\mathbf{U}$ is the transformation matrix, which also provides the necessary change of basis for other observables $\hat{O}$. This can be done by computing $\mathbf{U}^{-1}\mathbf{N}^{-1}\mathbf{O}\mathbf{U}$, in which the diagonal terms give exactly $\langle\phi_i|\hat{O}|\phi_i\rangle$.

\section{Results and discussion}
In this letter, for both $^4$He and $^4_\Lambda$H, we adopt interactions derived from leading-order pionless effective field theory (LO-$\slashed{\pi}$EFT), as outlined in the Supplementary Material. 
\subsection{Benchmark on the $^4$He system}
We first benchmark OP and NES methods on the $\alpha$ particle. To compute the excitation spectrum, we assign a dedicated neural-network branch to each state. Here, we adopt the network architecture introduced in our previous work \cite{ZHANG2026140285}. To balance expressivity and training efficiency, we employ a partial parameter-sharing scheme: all state branches share the same foundational MPNN, which establishes the underlying nodal structure of the many-body system, while each individual state is equipped with its own independent refining blocks, composed of multilayer perceptrons with residual connections. A short pretraining using the Adam optimizer with 4000 samples is applied to avoid inefficient initialization of the trial wavefunction, which is detailed in the Supplementary Material.

Within the original LO-$\slashed{\pi}$EFT framework, the $0^{+}_{2}$ state of $^{4}\mathrm{He}$ is a resonance in the $^{3}\mathrm{H} + p$ channel. 
Since our aim is to study bound-state behaviour, we retain only the two-body potential in the $^{4}\mathrm{He}$ calculations.
In Table \ref{Tab:He4_results}, we compare the performance of OP and NES methods in modeling the $0_1^+$ and $0_2^+$ states of $^4$He. Our results are benchmarked with the HH energy~\cite{PhysRevC.100.034004}, which is numerically exact for $s$-shell nuclei. The OP and NES results are highly consistent. Our VMC-NQS method underbinds both $0_1^+$ and $0_2^+$ states by $\sim 0.1$ MeV, with respect to the exact solution. Note that when employing a pure ground-state optimization, our NQS yields an energy of 
-39.84(0) MeV, in exact agreement with the HH benchmark. The small discrepancy in excited-state calculations stems from a variational trade-off, as the shared MPNN must balance the distinct spatial features of multiple states. This limitation could be systematically resolved in future work by increasing the network expressivity, for instance, by incorporating self-attention mechanisms~\cite{vonglehn2023selfattentionansatzabinitioquantum}. More importantly for spectroscopic studies, relative energies are of primary interest. Both OP and NES methods predict excitation energies in excellent agreement with the HH results, demonstrating robust systematic error cancellation.

Moreover, both OP and NES methods successfully capture the dramatic structural difference between the two states: while the ground state is highly compact with a point-proton radius of $r_p\approx 1.2$ fm, the first excited state exhibits a highly diffuse spatial distribution ($r_p\approx 3.1$ fm). This pronounced extension is characteristic of the 
$0_2^+$
state in 
$^4$He, which is widely interpreted as the monopole breathing mode.

In the OP calculations, the optimization weights are set to $w_1 = 1$ for the $0_1^+$ state and $w_2 = 0.5$ for the $0_2^+$ state.
Although the OP approach is flexible and allows straightforward evaluations of generic observables, it requires  tuning of the penalty strength $\lambda$.
We compare the performance for $\lambda = 30, 35$ and 500 MeV. As shown in Table \ref{Tab:He4_results}, $\lambda=35$ MeV yields an optimal balance.
When a smaller penalty strength ($\lambda=30$ MeV) is employed, the system suffers a slight variational collapse within the $J^\pi=0^+$ subspace, which consequently reduces the excitation energy. With a large penalty ($\lambda=500$ MeV),  the massive penalty term overwhelms the energy objective. To quickly drive the overlap to zero, the optimizer takes a shortcut by  mixing in spurious higher-angular-momentum components (as evidenced by $\langle \mathbf{L}^2 \rangle \approx 0.2$). Once the orthogonality is satisfied, the optimizer must then slowly navigate a stiff loss landscape to eliminate these unphysical modes. Consequently, the convergence becomes prohibitively slow. Within the same optimization steps where the $\lambda=35$ MeV case has  converged, the $\lambda=500$ MeV calculation remains trapped in this unphysical intermediate stage.

In contrast, the NES method obviates the need for hyperparameter tuning. Yielding results consistent with the optimal OP calculation, the NES converges to an excitation energy of $28.7$ MeV while maintaining the required $S$-wave symmetry.
\subsection{Excited-state structure and $M1$ transition of $^{4}_{\Lambda}\mathrm{H}$}
\begin{table*}[tbp]
    \centering
    \caption{\label{Tab:L4H_results} TABLE 2. Comparison of the ground state ($0^+$) and the first excited state ($1^+$) of $^{4}_{\Lambda}\mathrm{H}$ computed using the OP, OP-QNT and NES methods. The table presents the $\Lambda$ separation energy ($B_\Lambda$), distance between $\Lambda$ and the nuclear core ($r_{\Lambda N}$), expectation values of the squared orbital ($\langle \mathbf{L}^2 \rangle$) and spin ($\langle \mathbf{S}^2 \rangle$) angular momenta for each state, along with the excitation energy ($\Delta E$). The calculations are performed using the LO-$\slashed{\pi}$EFT Hamiltonian. The QNP results are taken from Ref.~\cite{ZHANG2026140285}.}
    \begin{tabular}{l c c c c c c c c c c}
        \toprule
        & & \multicolumn{4}{c}{Ground state ($0^+$)} & \multicolumn{4}{c}{First excited state ($1^+$)} & \\
        \cmidrule(lr){3-6} \cmidrule(lr){7-10} 
        Method & $\lambda$ [MeV] & $B_\Lambda$ [MeV] & $r_{\Lambda N}$ [fm] & $\langle \mathbf{L}^2 \rangle$ & $\langle \mathbf{S}^2 \rangle$ & $B_\Lambda$ [MeV] & $r_{\Lambda N}$ [fm] & $\langle \mathbf{L}^2 \rangle$ & $\langle \mathbf{S}^2 \rangle$ & $\Delta E$ [MeV] \\
        \midrule
        OP & 10 & 2.04(1) & 3.80(4) & 0.00(0) & 0.66(0) & 1.68(1) & 3.87(3) & 0.00(0) & 1.27(0) & 0.36(1) \\
        OP-QNT & 10 & 2.43(0) & 3.53(3) & 0.00(0) & 0.00(0) & 1.25(1) & 4.21(3) & 0.00(0) & 2.00(0) & 1.18(1) \\
        NES & -- & 2.43(1) & 3.58(3) & 0.00(0) & 0.00(0) & 1.24(1) & 4.29(4) & 0.00(0) & 2.00(0) & 1.19(1) \\
        QNP \cite{ZHANG2026140285} & -- & 2.44(0) & 3.51(1) & 0.00(0) & 0.00(0) & 1.26(0) & 4.34(1) & 0.00(0) & 2.00(0) & 1.18(0) \\
        \bottomrule
    \end{tabular}
\end{table*}
 Having benchmarked the accuracy of OP and NES methods on $^4$He, we now proceed to investigate the hypernucleus $^4_\Lambda$H. 
 Unlike the $0^+_2$ excited state in $^4$He, which shares the same quantum numbers as the ground state, the low-lying spectrum of $^4_\Lambda$H consists of a $0^+$ ground state and a $1^+$ excited state. This spin-doublet splitting is of profound interest as it directly constrains the spin-dependent components of the $YN$ interactions. 

 In our previous work~\cite{ZHANG2026140285}, we have investigated these two states of 
 $^{4}_{\Lambda}\mathrm{H}$ by constructing trial wavefunctions 
 with fixed quantum numbers using projection operators, 
 which we dubbed QNP. Given the $\sim$1.2 MeV excitation energy of $^{4}_{\Lambda}\mathrm{H}$, setting $\lambda = 10$ MeV can effectively prevent variational collapse in OP method.
However, accurately extracting these low-lying hypernuclear states within standard VMC-NQS framework is hindered by spin contamination, which is  driven by the near-degeneracy of adjacent spin states~\cite{ZHANG2026140285}. With energy splittings smaller than the $B_\Lambda$, the unconstrained optimization easily gets trapped in unphysical local minima.  As shown in the first row of Table \ref{Tab:L4H_results}, when solely the energy objective and overlap penalty is minimized, the optimizer struggles to disentangle the $0^+$ and $1^+$ states. To eradicate this spurious mixing, we propose the quantum number targeting (QNT) technique. That is, we augment the loss function with an explicit penalty for the total spin-square operator, $\mathbf{S}^2$
\begin{align}
	L \rightarrow L +\lambda_s\sum_{i=1}^K w_i(S[\Psi_i] -s_i(s_i+1))^2, 
\end{align}
where 
$S[\Psi_i] = \langle\Psi_i|\mathbf{S}^2|\Psi_i\rangle$ ,$s_i$ is the target spin number for state $\Psi_i$ and $\lambda_s=3$ MeV is the corresponding weight. A detailed analysis of the dependence on 
$\lambda_s$ is provided in the Supplementary Material.

Panel (a) of Fig.~\ref{Fig:QNT} shows the optimization process of $^{4}_{\Lambda}\mathrm{H}$ with and without QNT. As shown in Fig.~\ref{Fig:QNT} (b), while both optimizations successfully drive the absolute overlap $S_{12}$ toward zero, only the QNT-assisted optimization correctly reproduces the QNP energy. This indicates that the standard OP method merely converges to two mutually orthogonal superpositions of the $0^{+}$ and $1^{+}$ states, which correspond to suboptimal local minima of the loss landscape. 
 \begin{figure}[tb]
    \centering
    \includegraphics[width=\linewidth]{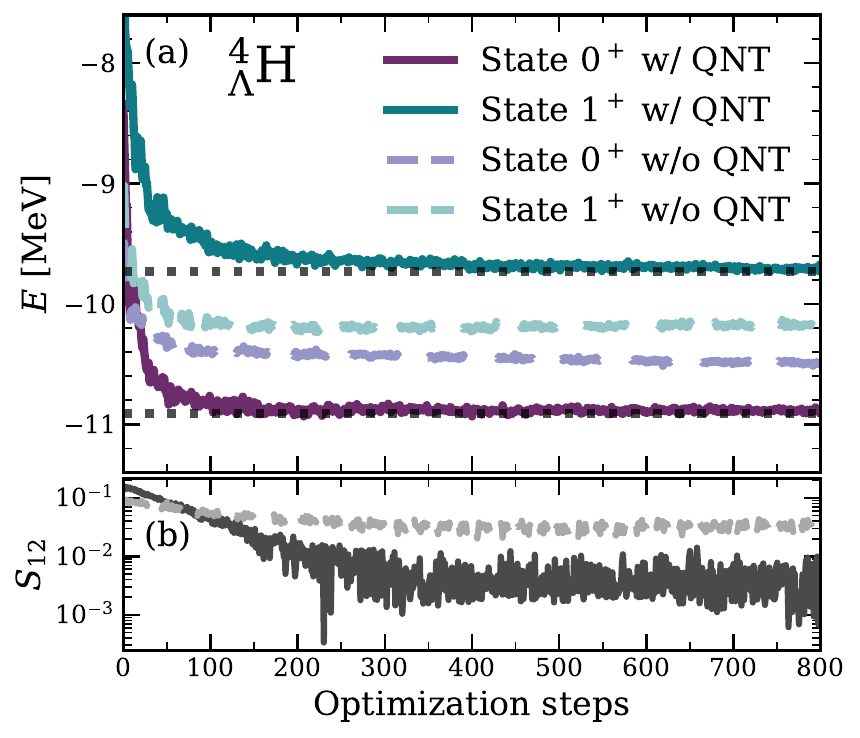}
    \caption{Fig. 1. Comparison of optimization for $^{4}_{\Lambda}\mathrm{H}$ in the overlap penalty method (OP) with and without the quantum number targeting (QNT) technique. (a) Energy convergence for the $0^+$ ground state (purple) and $1^+$ excited state (cyan). Black dotted lines represent the reference energies taken from Ref.~\cite{ZHANG2026140285}. (b) Absolute overlap $S_{12}$ between the two states. The dark grey solid line represents the optimization with QNT, while the light grey dashed line denotes that without QNT.}
    \label{Fig:QNT}
\end{figure}
\begin{figure}[tb]
    \centering
    \includegraphics[width=\linewidth]{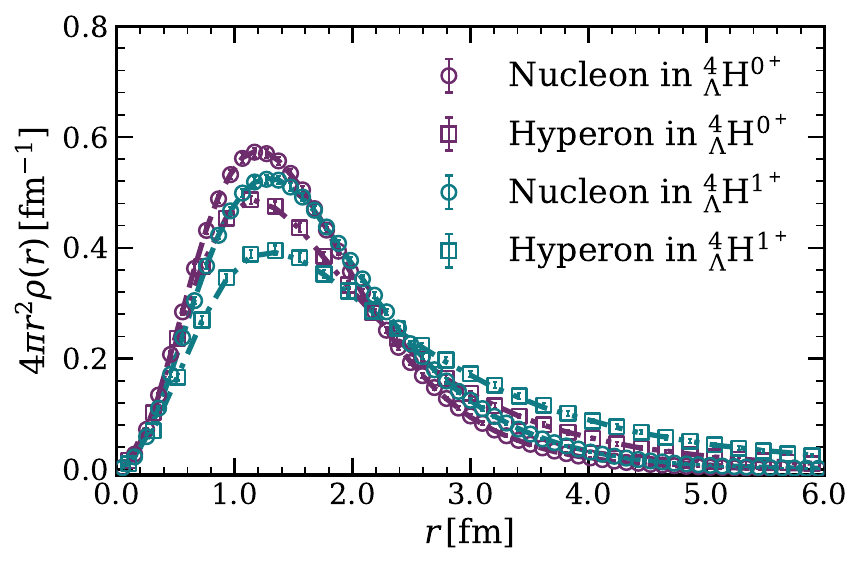}
    \caption{Fig. 2. Radial density distributions $4\pi r^2 \rho(r)$ of nucleons and the $\Lambda$ hyperon for the ground ($0^+$) and excited ($1^+$) states of the $^{4}_{\Lambda}\mathrm{H}$ hypernucleus. The purple and teal colors denote the $0^+$ and $1^+$ states, respectively. The open symbols with error bars represent the results obtained from the overlap penalty method with the quantum number targeting technique, where circles correspond to nucleons and squares to the hyperon. The dashed (nucleon) and dash-dotted (hyperon) lines represent the distributions calculated using the quantum number projection method~\cite{ZHANG2026140285}.}
    \label{Fig:distribution}
\end{figure}
Table \ref{Tab:L4H_results} summarizes the converged properties of the $0^+$ and $1^+$ states in $^4_\Lambda$H using the OP-QNT, NES, and the rigorous QNP methods. The QNP method, which intrinsically projects out the exact spin components at the cost of significantly higher computational overhead, serves as a stringent baseline in our study. Remarkably, both the OP-QNT and the  NES methods yield results in excellent agreement with the  QNP calculations. We obtain a ground-state separation energy $B_\Lambda = 2.43$ MeV and an excited-state separation energy $B_\Lambda = 1.24$ MeV. Furthermore, all methods give a highly consistent distance between the $\Lambda$ hyperon and the nuclear core ($r_{\Lambda N}$). Notably, because NES recovers the true eigenstates by diagonalizing the Hamiltonian within the optimized subspace rather than forcing strict orthogonality during training, it effectively circumvents the spin contamination  in the unconstrained OP approach.

Beyond the energy spectrum, understanding the spatial structure of hypernuclei is crucial for modeling their production and decay mechanisms~\cite{Chen:2024eaq,Chen:2025eeb,Sun:2025oib,Wang2025}. As the observable extraction in the NES method requires further orthogonalization of the states, we use the OP-QNT method to compute the spatial distribution for convenience. Fig. \ref{Fig:distribution} presents the radial density distributions of nucleons and the $\Lambda$ hyperon for both the $0^+$ and $1^+$ states, which is defined as
\begin{align}
\rho_N(r) &= \frac{1}{4\pi r^2} \frac{\langle\Psi|\sum_i \delta(\|\bm{r}_i - \bm{R}_{\text{c.m.}}\| - r)|\Psi\rangle}{(A-1)\langle\Psi|\Psi\rangle}, \nonumber\\
\rho_\Lambda(r) &= \frac{1}{4\pi r^2} \frac{\langle\Psi|\delta(\|\bm{r}_\Lambda - \bm{R}_{\text{c.m.}}\| - r)|\Psi\rangle}{\langle\Psi|\Psi\rangle},
\end{align} 
where $\bm{R}_{\text{c.m.}}$ is the center-of-mass coordinate of the hypernuclei. 
In Fig. \ref{Fig:distribution}, the open symbols represent the distributions extracted using the OP-QNT method, which perfectly match the QNP results depicted by the dashed and dash-dotted lines. Comparing the two states, the hyperon in the $1^+$ excited state exhibits a broader and more diffuse density profile than in the $0^+$ state.  These results once again corroborate the accuracy of OP and NES methods in modeling excited-state properties of nuclei and hypernuclei.

Beyond the above diagonal properties, evaluating off-diagonal observables like the $M1$ $\gamma$ transition is also of great interest. For the transition from the $1^+$ excited state to the $0^+$ ground state, the reduced transition probability $B(M1)$ is defined as~\cite{Dalitz1978}
\begin{equation}
    B(M1; 1^+ \rightarrow 0^+) = \frac{1}{2J_i + 1} \left| \langle \Psi^{0^+} || O(M1) || \Psi^{1^+} \rangle \right|^2,
\end{equation}
where $J_i = 1$ is the total spin of the initial state. Incorporating the standard geometric factor, the magnetic dipole operator is written as $O(M1) = \sqrt{3/4\pi} \, \boldsymbol{\mu}$. $\bm{\mu}$ is the magnetic moment
\begin{align}
    \boldsymbol{\mu} = \sum_{i=1}^{A-1} \left( \mathcal{P}_i^{p} \mathbf{l}_i + \mu_{p} \mathcal{P}_i^{p}\boldsymbol{\sigma}_i + \mu_{n} \mathcal{P}_i^{n}\boldsymbol{\sigma}_i\right) + \mu_\Lambda \boldsymbol{\sigma}_\Lambda,
\end{align}
where $\mathcal{P}^{p(n)}_i$ is the proton (neutron) projection operator and $\mu_p=2.793$ $\mu_N$, $\mu_n=-1.913$ $\mu_N$, $\mu_\Lambda=-0.613$ $\mu_N$ and $\mu_N$ the nuclear magneton.
Since the $^{4}_{\Lambda}\mathrm{H}$ hypernucleus is  dominated by the $S$-wave ($L=0$) configuration, the orbital angular momentum contribution vanishes. Consequently, this $\gamma$ transition is purely governed by the spin magnetization of baryons.
In the  weak-coupling limit (WCL)~\cite{Dalitz1978, RevModPhys.88.035004}, the $^{3}\mathrm{H}$ core and the $\Lambda$ hyperon are assumed to be weakly coupled without any structural modification. Under this assumption, the spatial wavefunctions of the initial and final states are identical, and the transition strength is solely determined by the spin configuration. For $^4_\Lambda$H, the WCL prediction reads
\begin{align}
    B(M1; 1^+ \rightarrow 0^+)_{\mathrm{WCL}} = \frac{3}{4\pi} (\mu_{^3\mathrm{H}} - \mu_\Lambda)^2,
    \label{eq:BM1_weak_coupling}
\end{align}
where $\mu_{^3\text{H}}$ is the magnetic moment of $^3$H.
Our VMC-NQS calculation gives $\mu_{^3\text{H}} = 2.73(0)$ $\mu_N$, which leads to $B(M1; 1^+ \rightarrow 0^+)_{\text{WCL}}=2.668$ $\mu_N^2$.

Over the past decades, microscopic methods such as  Faddeev-Yakubovsky~\cite{Faddev}, VMC~\cite{Usmani_2008,PhysRevC.73.011302}, AFDMC~\cite{LONARDONI2013243,PhysRevC.89.014314}, SVM~\cite{Nemura1999StudyOL, PhysRevC.106.L031001, SVMPRL, Contessi2019csf, Schafer2020rba, Schfer2021ConsequencesOI, newSVM1}, no-core shell model~\cite{newNCSM1, newNCSM2, PRLLe, Le2025}, Gomow shell model~\cite{LI2025139708} and nuclear lattice effective field theory~\cite{Frame2020ImpurityLM, Hildenbrand2022,Hildenbrand2024, TONG2025825, tong2025hyperneutronstarsabinitio, tong2025multistrangenessmatterabinitio} have achieved remarkable successes in describing the diagonal properties of hypernuclei. However, to the best of our knowledge, \textit{ab initio} calculation of the $M1$ transition rate for $^{4}_{\Lambda}\mathrm{H}$ has not been reported previously, leaving the WCL as the sole reference. 
As summarized in Table~\ref{Tab:M1_results},  OP-QNT and NES calculations yield highly consistent results within statistical error. Our predicted $B(M1)$ values exhibit a marginal suppression of  $\sim$1.3\% relative to the WCL value. Unlike light $p$-shell hypernuclei where the \( \Lambda \)-hyperon severely distorts the core and breaks the weak-coupling picture~\cite{PhysRevC.59.2351}, this agreement highlights the exceptional rigidity of the $^3$H core. The stronger attraction in the spin-singlet channel makes the $\Lambda$-$^3$H spatial wavefunction of the $0^+$ state more compact than that of the $1^+$ state, as seen in Fig. \ref{Fig:distribution}. This difference in the relative spatial wavefunctions leads to an incomplete overlap ($< 1$) during the transition, which naturally explains the tiny suppression.
\begin{table}[tb]
    \centering
    \caption{\label{Tab:M1_results} TABLE 3. The reduced magnetic dipole transition probability $B(M1)$ (in units of $\mu_N^2$) for the $1^+ \rightarrow 0^+$ transition in $^{4}_{\Lambda}\mathrm{H}$. Results obtained from the OP-QNT and NES methods are compared with the theoretical weak-coupling limit.}
    \begin{tabular}{l c}
        \toprule
        Method & $B(M1; 1^+ \rightarrow 0^+)$ $[\mu_N^2]$ \\
        \midrule
        WCL & $2.668(0)$ \\
        OP-QNT  & $2.634(5)$ \\  
        NES & $2.630(7)$ \\ 
        \bottomrule
    \end{tabular}
\end{table}

\section{Summary and outlook}
\label{sum}
In summary, we have 
established the first excited-state framework for nuclear and hypernuclear NQS calculations.
By systematically comparing the OP and the NES methods, we evaluate their performance on the $A=4$ systems. We demonstrated that while the OP method requires careful tuning of the penalty strength to avoid unphysical local minima, the NES method serves as a robust, hyperparameter-free alternative that intrinsically prevents variational collapse.

When extending the framework to the $^{4}_{\Lambda}\mathrm{H}$ hypernucleus, we identify a severe spin contamination issue in the unconstrained OP approach, similar to the case in the ground-state calculations. To overcome this difficulty, we introduce the QNT technique to impose spin symmetry constraints. The NES method mitigates spin contamination by recovering the true eigenstates via diagonalization within the optimized subspace. Both OP-QNT and NES results are in excellent agreement with the QNP calculations, accurately reproducing the binding and excitation energies as well as the spatial structures. We further evaluated the off-diagonal $M1$ $\gamma$ transition for $^{4}_{\Lambda}\mathrm{H}$. The calculated $B(M1)$ is suppressed by  $\sim$1.3\%  relative to the WCL, reflecting the rigidity of the  \( ^3\text{H} \) core. This small reduction stems from incomplete spatial overlap induced by the spin-dependent $YN$ interaction.

The accurate excited-state algorithms developed here lay the groundwork for systematic \textit{ab initio} calculations of heavier nuclear and hypernuclear systems. An immediate application is to extend this framework to the $A=4$ hypernuclear isodoublet ($^{4}_{\Lambda}\mathrm{H}$ and $^{4}_{\Lambda}\mathrm{He}$) to investigate charge symmetry breaking (CSB) effects, which will offer tighter constraints on the spin-dependent $YN$  forces. Furthermore, by incorporating  chiral effective field theory interactions, the VMC-NQS approach will serve as a highly predictive and versatile tool for exploring low-energy nuclear and hypernuclear spectroscopy.

\section*{Acknowledgments}
\label{acknowledgments}
 This work is partially supported by the National Natural Science Foundation of China under Contracts No. 12475187 and 124B1006,
 the National Key Research and Development Project of China under Grant No.  2024YFA1610702.
 The computations in this research were performed using the CFFF platform of Fudan University.

\bibliographystyle{elsarticle-num-names}
\bibliography{rif}

\end{document}


\title{Supplementary Material for ``Neural-network excited states of $A=4$ nuclei and hypernuclei"}
\author{Zi-Xiao Zhang}
\affiliation{Key Laboratory of Nuclear Physics and Ion-beam Application (MOE), Institute of Modern Physics, Fudan University, Shanghai 200433, China}

\author{Yi-Long Yang}
\affiliation{State Key Laboratory of Nuclear Physics and Technology, School of Physics, Peking University, Beijing 100871, China}

\author{Xiao-Lu Qian}
\affiliation{Department of Physics, Fudan University, Shanghai 200433, China}

\author{Wan-Bing He}
\email{hewanbing@fudan.edu.cn}
\affiliation{Key Laboratory of Nuclear Physics and Ion-beam Application (MOE), Institute of Modern Physics, Fudan University, Shanghai 200433, China}
\affiliation{ Shanghai Research Center for Theoretical Nuclear Physics, NSFC and Fudan University, Shanghai 200438, China}

\author{Peng-Wei Zhao}
\affiliation{State Key Laboratory of Nuclear Physics and Technology, School of Physics, Peking University, Beijing 100871, China}

\author{Bing-Nan Lu}
\affiliation{Graduate School of China Academy of Engineering Physics, Beijing 100193, China}

\author{Yu-Gang Ma}
\email{mayugang@fudan.edu.cn}
\affiliation{Key Laboratory of Nuclear Physics and Ion-beam Application (MOE), Institute of Modern Physics, Fudan University, Shanghai 200433, China}
\affiliation{ Shanghai Research Center for Theoretical Nuclear Physics, NSFC and Fudan University, Shanghai 200438, China}
\affiliation{School of Physics, East China Normal University, 200062, Shanghai, China}

\begin{abstract}
This supplementary material provides details on the Hamiltonian, pretraining protocol, dependence on  quantum number targeting strength $\lambda_s$, Monte Carlo estimation for the gradients of overlap between states, the alternative proof for the objective functional in natural excited states method and Cartesian representation of  $\mathbf{L}^2$ operator. 
\end{abstract}

\maketitle

\section{Hamiltonian}
In this work, we adopt the Hamiltonian derived within leading-order (LO) pionless effective field theory ($\slashed{\pi}$EFT) for both atomic nuclei and hypernuclei. For benchmark on $^4$He, to facilitate comparision with hyperspherical harmonics method (HH), we adopt the identical two-body force as used in Ref~\cite{PhysRevC.100.034004}, which reads
 \begin{align}
v_{ij} = \left[V_{10} \mathcal{P}_{10}(ij)  + V_{01} \mathcal{P}_{01}(ij)\right]e^{-(r_{ij}/r_0)^2} .
 \end{align}
In the above equation, $V_{10}=-60.575$ MeV, $V_{01}=-37.9$ MeV and $r_0$ is set to 1.65 fm. $\mathcal{P}_{01}(ij)$ and $\mathcal{P}_{10}(ij)$ are the spin-isospin projection operators on $S/T=0/1$ and $1/0$ channels, respectively. 

In the hypernuclear calculations, we use the model \enquote{o} in Ref~\cite{ZHANG2026140285}.
The interaction among nucleons is taken as the model \enquote{o} in Ref~\cite{PhysRevC.103.054003}
, with the nucleon-nucleon potential greatly reproduces the $np$ scattering lengths
 \begin{align}
v_{ij} = C_{10} \mathcal{P}_{10}(ij) C_1(r_{ij}) + C_{01} \mathcal{P}_{01}(ij) C_0(r_{ij}).
 \end{align}
 in which $C_1(r_{ij})$ and $C_0(r_{ij})$ are the corresponding Gaussian regulators, fitted separately for each isospin channel. 
 To reproduce the binding energies for $A\geq3$ nuclei and avoid \enquote{Thomas collapse}, a three-body repulsive force is introduced, which is of the form
 \begin{align}
 	V_{ijk} = \frac{c_E}{f_{\pi}^4 \Lambda_\chi} \frac{\left(\hbar c\right)^6}{\pi^3 R_3^6} \sum_{\text{cyc}} e^{-\left(r_{ij}^2 + r_{jk}^2\right)/R_3^2}.
 \end{align}
In the above, $\Lambda_\chi=1$ GeV, $f_{\pi}=92.4$ MeV is the pion decay constant. The $c_E$ is fixed to reproduce the $^3$H binding energy, $B(^3\text{H})=8.475$ MeV, for a given value of the cutoff $R_3$.

For the $YN$ interaction, we use $s$-wave $\Lambda N$ and $\Lambda NN$ contact interactions modeled by LO-$\slashed{\pi}$EFT
\begin{align}
	V_{\Lambda N} &= \sum_S C_\lambda^S \sum_i \mathcal{P}_S(\Lambda i)e^{-\frac{\lambda^2}{4}r_{\Lambda i}^2},\nonumber\\
	V_{\Lambda NN} &= \sum_{SI} D_\lambda^{SI} \sum_{i<j} \mathcal{Q}_{SI}(\Lambda ij)e^{-\frac{\lambda^2}{4}(r_{\Lambda i}^2 + r_{\Lambda j}^2)},
\end{align}
with original low-energy constants  sourced from Ref~\cite{PhysRevC.106.L031001}. $\mathcal{P}_S(\Lambda i)$ and $\mathcal{Q}_{SI}(\Lambda ij)$ are projection operators acting on $\Lambda N$ pairs (spin $S$) and $\Lambda NN$ triplets (spin $S$ and isospin $I$), respectively. In this work, we set the momentum cutoff $\lambda$ to be 2 fm$^{-1}$. As shown in Ref. \cite{ZHANG2026140285}, this hypernuclear model can reproduce the experimental $\Lambda$ separation energy, $B_\Lambda$ values, for $A\leq 13$ charge-symmetric hypernuclei. Equipped with this rigorously constrained Hamiltonian, we  proceed to investigate the excited-state properties of the 
$A=4$ systems.

\section{Pretraining}
    \begin{figure}[tb]
	\centering
	\subfigure
	{
		\begin{minipage}[]{1\linewidth}
			\centering
			\includegraphics[scale=0.75]{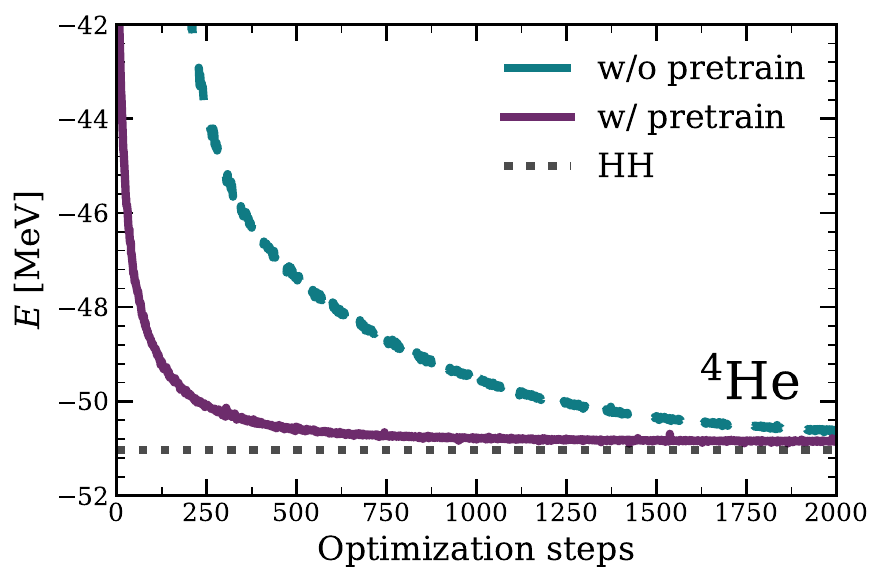}
		\end{minipage}
	}
	
	\caption{
		Convergence of the total energy for 
		$^4$He computed using the natural excited states method as a function of optimization steps. The solid purple curve and the dashed teal curve represent the training trajectories with and without pretraining. The dotted gray line denotes the results from hyperspherical harmonics method (HH)~\cite{PhysRevC.100.034004}.
	}
	\label{Fig:pretrain}
\end{figure}

A naive random initialization of the ansatz typically produces unphysical starting points, rendering the initial phase of the VMC optimization highly inefficient. To overcome this cold-start problem, we employ a pretraining strategy. Specifically, we warm up the network parameters using the Adam optimizer~\cite{kingma2015adam} with a small batch of $4000$ samples for a short period. This lightweight pretraining efficiently steers the wave functions into a physically reasonable region of the parameter space, thereby drastically reducing the computational cost required for the full energy optimization.

To demonstrate the effectiveness of our pretraining strategy, Fig.~\ref{Fig:pretrain} compares the energy convergence trajectories of $^{4}\mathrm{He}$ with and without the initial warm-up phase, computed with the NES method. The substantial disparity between the two curves highlights the challenge of navigating the highly non-linear loss landscape from a random starting point. The pretraining protocol provides a crucial initial push, rapidly guiding the optimization into the physical basin of attraction. This pretraining strategy is therefore adopted for all  calculations in this work.
\section{Dependence on  quantum number targeting strength $\lambda_s$}
In this section, we investigate the influence of the quantum number targeting (QNT) weight $\lambda_s$ on the physical properties of the $^4_\Lambda$H states to justify our choice of $\lambda_s = 3$ MeV in the main text. 
As shown in Table \ref{Tab:L4H_results}, the introduction of the QNT penalty effectively suppresses spin contamination. For a small weight of $\lambda_s = 1$ MeV, a slight residual spin contamination remains ($\langle \mathbf{S}^2 \rangle = 0.01$ and $1.99$ for the $0^+$ and $1^+$ states, respectively). Once $\lambda_s \ge 3$ MeV, the expectation values of $\langle \mathbf{S}^2 \rangle$ are strictly locked to their exact target values of 0 and 2. Notably, the separation energies $B_\Lambda$ remain remarkably stable across all tested $\lambda_s$ values, showing excellent agreement with the quantum number projection (QNP) benchmarks~\cite{ZHANG2026140285}.

However, a closer inspection of the spatial structure ($r_{\Lambda N}$) reveals a critical pathology associated with overly strong penalty terms in weakly bound systems. While the energy is insensitive to large $\lambda_s$, the wavefunction undergoes an unphysical spatial expansion. For instance, when $\lambda_s$ reaches 50 MeV, $r_{\Lambda N}$ for the ground state abruptly increases to 4.38(4) fm. Similarly, at $\lambda_s = 100$ MeV, the excited-state radius expands to 4.90(4) fm. 
This phenomenon originates from the shallow energy landscape of the weakly bound $\Lambda$ hyperon. The spin contamination is primarily driven by the hyperon-nucleon ($YN$) interactions. To trivially minimize a heavily weighted spin penalty without losing significant binding energy, the optimizer tends to push the $\Lambda$-hyperon into the asymptotic region. In this dilute limit, the $YN$ interaction vanishes, making it mathematically easier to construct a pure spin state. The kinetic energy reduction and potential energy loss cancel out, leaving the total energy seemingly correct, but at the cost of a severely distorted, overly diffuse wavefunction.

Based on the results in Table~\ref{Tab:L4H_results}, we identify $\lambda_s = 3$ MeV as the optimal ``sweet spot.'' At this value, the QNT method successfully eradicates spin contamination while preserving the correct spatial compactness ($r_{\Lambda N} \approx 3.53$ fm for $0^+$ and $4.21$ fm for $1^+$), perfectly consistent with the QNP calculations. Ensuring this correct spatial structure is of great importance for the reliable evaluation of structure-sensitive observables, such as the  $M1$ transitions discussed in the main text.
\begin{table*}[htbp]
	\caption{\label{Tab:L4H_results} Dependence of the calculated properties for the $^4_\Lambda$H ground ($0^+$) and first excited ($1^+$) states on the quantum number targeting weight $\lambda_s$ within the OP-QNT method. The table lists the $\Lambda$ separation energy ($B_\Lambda$), the distance between the $\Lambda$ hyperon and the nuclear core ($r_{\Lambda N}$), and the expectation values of the squared orbital ($\langle \mathbf{L}^2 \rangle$) and spin ($\langle \mathbf{S}^2 \rangle$) angular momenta. Results from the quantum number projection (QNP) approach are provided as a benchmark~\cite{ZHANG2026140285}.  }
	\begin{ruledtabular} 
		\begin{tabular}{l c c c c c c c c c c}
			
			& & \multicolumn{4}{c}{Ground state ($0^+$)} & \multicolumn{4}{c}{First excited state ($1^+$)} & \\
			\cline{3-6} \cline{7-10} 
			
			\rule{0pt}{3ex}Method & $\lambda_s$ [MeV] & $B_\Lambda$ [MeV] & $r_{\Lambda N}$ [fm] & $\langle \mathbf{L}^2 \rangle$ & $\langle \mathbf{S}^2 \rangle$ & $B_\Lambda$ [MeV] & $r_{\Lambda N}$ [fm] & $\langle \mathbf{L}^2 \rangle$ & $\langle \mathbf{S}^2 \rangle$ &  \\
			\colrule 
			\rule{0pt}{3ex}OP-QNT & 1 & 2.42(1) & 3.53(2) & 0.00(0) & 0.01(0) & 1.25(1)  & 4.15(4) & 0.00(0) & 1.99(0)\\
			 & 3 & 2.43(0) & 3.53(3) & 0.00(0) & 0.00(0) & 1.25(1)  & 4.21(3) & 0.00(0) & 2.00(0)  \\
		 & 10 & 2.43(1) & 3.54(3) & 0.00(0) & 0.00(0) & 1.25(1)  & 4.11(3) & 0.00(0) & 2.00(0) \\
			 & 50 & 2.43(0) & 4.38(4) & 0.00(0) & 0.00(0) &1.25(1)  & 4.30(3) & 0.00(0) &  2.00(0) \\
			 & 100 & 2.43(1) & 3.67(3) & 0.00(0) & 0.00(0) &1.23(1)  & 4.90(4) & 0.00(0) &  2.00(0)  \\
			QNP \cite{ZHANG2026140285} & -- & 2.44(0) & 3.51(1) & 0.00(0) & 0.00(0) & 1.26(0) & 4.34(1) & 0.00(0) &  2.00(0)  \\
		\end{tabular}
	\end{ruledtabular} 
\end{table*}
\section{Monte Carlo gradient estimation of overlap penalty}
\label{overlap penalty}
 We present the detailed derivation for the gradient of the overlap penalty term. 
 The absolute squared overlap between two states $|\Psi_i\rangle$ and $|\Psi_j\rangle$ is defined as
 \begin{equation}
 	|S_{ij}|^2 = \frac{\langle \Psi_i | \Psi_j \rangle \langle \Psi_j | \Psi_i \rangle}{\langle \Psi_i | \Psi_i \rangle \langle \Psi_j | \Psi_j \rangle}.
 \end{equation}
 To evaluate this efficiently with Monte Carlo, we introduce the integrated overlap ratio $\bar{O}_{ij}$ and the local overlap $O_{ij}(\mathbf{x})$
 \begin{equation}
 	\bar{O}_{ij} = \frac{\langle \Psi_i | \Psi_j \rangle}{\langle \Psi_i | \Psi_i \rangle}, \quad \quad O_{ij}(\mathbf{x}) = \frac{\Psi_j(\mathbf{x})}{\Psi_i(\mathbf{x})},
 \end{equation}
 where $\mathbf{x}$ denotes the set of single-nucleon spatial three-dimensional coordinates and the z-projection of the spin-isospin degrees of freedom.
 Note that the overlap penalty can be factored as $|S_{ij}|^2 = \bar{O}_{ij} \bar{O}_{ji}$.
 
 Next, we compute the gradient of $|S_{ij}|^2$ with respect to variational parameter $\theta$ belonging to the network $\Psi_i$. Using the quotient rule, we collect all terms involving the derivative $\partial_\theta \Psi_i$, denoted as $A_{ij} \triangleq \partial_{\theta \in \Psi_i} |S_{ij}|^2$
 \begin{align}
 	A_{ij} &= \frac{\langle \partial_\theta \Psi_i | \Psi_j \rangle \langle \Psi_j | \Psi_i \rangle + \langle \Psi_i | \Psi_j \rangle \langle \Psi_j | \partial_\theta \Psi_i \rangle}{\langle \Psi_i | \Psi_i \rangle \langle \Psi_j | \Psi_j \rangle} \nonumber \\
 	&- |S_{ij}|^2 \frac{\langle \partial_\theta \Psi_i | \Psi_i \rangle \langle \Psi_j | \Psi_j \rangle + \langle \Psi_i | \partial_\theta \Psi_i \rangle \langle \Psi_j | \Psi_j \rangle}{\langle \Psi_i | \Psi_i \rangle \langle \Psi_j | \Psi_j \rangle}.
 \end{align}
 
 Because $|S_{ij}|^2$ is real, we can simplify the expression using the complex conjugate relation. The first term can be rewritten by recognizing that $\langle \Psi_j | \Psi_i \rangle / \langle \Psi_j | \Psi_j \rangle = \bar{O}_{ji}$ 
 \begin{align}
 	\bar{O}_{ji} \frac{\langle \partial_\theta \Psi_i | \Psi_j \rangle}{\langle \Psi_i | \Psi_i \rangle} + \bar{O}_{ji}^* \frac{\langle \Psi_j | \partial_\theta \Psi_i \rangle}{\langle \Psi_i | \Psi_i \rangle} = 2 \text{Re} \left[ \bar{O}_{ji}^* \frac{\langle \Psi_j | \partial_\theta \Psi_i \rangle}{\langle \Psi_i | \Psi_i \rangle} \right].
 \end{align}
 Similarly, the subtraction term simplifies to
 \begin{equation}
 	- 2 \text{Re} \left[ |S_{ij}|^2 \frac{\langle \Psi_i | \partial_\theta \Psi_i \rangle}{\langle \Psi_i | \Psi_i \rangle} \right].
 \end{equation}
 Combining these and using the relation $|S_{ij}|^2 = \bar{O}_{ij}^* \bar{O}_{ji}^*$, we obtain
 \begin{equation}
 	A_{ij} = 2 \text{Re} \left[ \bar{O}_{ji}^* \frac{\langle \Psi_j | \partial_\theta \Psi_i \rangle}{\langle \Psi_i | \Psi_i \rangle} - \bar{O}_{ij}^* \bar{O}_{ji}^* \frac{\langle \Psi_i | \partial_\theta \Psi_i \rangle}{\langle \Psi_i | \Psi_i \rangle} \right].
 	\label{eq:Aij_analytical}
 \end{equation}
 
 To cast Eq.~(\ref{eq:Aij_analytical}) into a form suitable for Monte Carlo sampling, we employ the log-derivative trick, decomposing the integrals into expectations over the probability density $P(\mathbf{x}) \propto |\Psi_i(\mathbf{x})|^2$
 \begin{align}
 	\frac{\langle \Psi_j | \partial_\theta \Psi_i \rangle}{\langle \Psi_i | \Psi_i \rangle} &= \mathbb{E}_i \left[ O_{ij}^*(\mathbf{x}) \partial_\theta \ln \Psi_i(\mathbf{x}) \right], \label{eq:mc1} \\
 	\frac{\langle \Psi_i | \partial_\theta \Psi_i \rangle}{\langle \Psi_i | \Psi_i \rangle} &= \mathbb{E}_i \left[ \partial_\theta \ln \Psi_i(\mathbf{x}) \right]. \label{eq:mc2}
 \end{align}
 Substituting Eq.~(\ref{eq:mc1}) and Eq.~(\ref{eq:mc2}) back into the total expression, we arrive at the Monte Carlo estimator for the gradient
 \begin{equation}
 	A_{ij} = 2 \text{Re} \left[ \bar{O}_{ji}^* \left( \mathbb{E}_i \left[ O_{ij}^* \partial_\theta \ln \Psi_i \right] - \bar{O}_{ij}^* \mathbb{E}_i \left[ \partial_\theta \ln \Psi_i \right] \right) \right].
 	\label{eq:overlap_gradient_final}
 \end{equation}
 In a sequential optimization scheme (where $\Psi_j$ is fixed and $\Psi_i$ is optimized), the parameter update is governed solely by $A_{ij}$. If both states are updated simultaneously and share parameters, the total derivative is given by the symmetric sum $A_{ij} + A_{ji}$.
 \section{Alternative Derivation of the Natural Excited States Method Objective Using Jacobi’s Formula}
\label{Jacobi}
In this section, we provide a concise alternative derivation of the natural excited states method objective function [Eq. (7) in the main text] using  Jacobi's formula~\cite{vein1999determinants}.

Recall that the overlap matrix $\mathbf{N}$ and the total Hamiltonian matrix $\mathbf{H}$ are defined by their elements $\mathbf{N}_{ij} = \langle \psi_i | \psi_j \rangle$ and $\mathbf{H}_{ij} = \langle \psi_i | \hat{H} | \psi_j \rangle$, respectively. We introduce an parameter-dependent matrix $\mathbf{M}(\epsilon)$ defined by the expectation of an exponential operator
\begin{equation}
	\mathbf{M}_{ij}(\epsilon) = \langle \psi_i | e^{\epsilon \hat{H}} | \psi_j \rangle.
\end{equation}
The first derivative of the determinant of $\mathbf{M}(\epsilon)$ evaluated at $\epsilon = 0$ is directly related to the sum of the Hamiltonian expectation values 
\begin{equation}
	\left. \frac{\partial}{\partial \epsilon} \det \mathbf{M}(\epsilon) \right|_{\epsilon=0} = \det(\mathbf{N}) \frac{\langle \Psi | \hat{\mathcal{H}} | \Psi \rangle}{\langle \Psi | \Psi \rangle},
	\label{eq:app_derivative_def}
\end{equation}
where $\Psi$ represents the $K$-state Slater determinant, and $\hat{\mathcal{H}} = \hat{H}_1\oplus...\oplus\hat{H}_K$.

Jacobi's formula relates the derivative of a matrix determinant to its trace
\begin{equation}
	\frac{\partial}{\partial \epsilon} \det \mathbf{M}(\epsilon) = \det \mathbf{M}(\epsilon) \text{Tr} \left( \mathbf{M}(\epsilon)^{-1} \frac{\partial \mathbf{M}(\epsilon)}{\partial \epsilon} \right).
	\label{eq:app_jacobi}
\end{equation}
Evaluating the terms in Eq.~(\ref{eq:app_jacobi}) at $\epsilon = 0$, we have $\mathbf{M}(0)=\mathbf{N}$, $ \frac{\partial \mathbf{M}(\epsilon)}{\partial \epsilon} |_{\epsilon=0} = \mathbf{H}$.
Substituting these relations back into Jacobi's formula gives
\begin{equation}
	\left. \frac{\partial}{\partial \epsilon} \det \mathbf{M}(\epsilon) \right|_{\epsilon=0} = \det(\mathbf{N}) \text{Tr} \left[ \mathbf{N}^{-1} \mathbf{H} \right].
	\label{eq:app_jacobi_apply}
\end{equation}
By equating Eq.~(\ref{eq:app_derivative_def}) and Eq.~(\ref{eq:app_jacobi_apply}), and canceling out $\det(\mathbf{N})$, we directly obtain the NES loss function
\begin{equation}
	\frac{\langle \Psi | \hat{\mathcal{H}} | \Psi \rangle}{\langle \Psi | \Psi \rangle} = \text{Tr} \left[ \mathbf{N}^{-1} \mathbf{H} \right].
\end{equation}
\section{Cartesian representation of  $\mathbf{L}^2$ operator}
To facilitate numerical implementations in VMC framework,  we provide the explicit differential forms of the total orbital angular momentum operator $\mathbf{L}^2$. The operator acting on the wavefunction $\Psi(\mathbf{x})$ can be decomposed into one-body and two-body contributions as follows
\begin{align}
\mathbf{L}^2 \Psi(\mathbf{x}) &= \left( \sum_{i=1}^A \boldsymbol{l}_i^2 + 2 \sum_{i<j} \boldsymbol{l}_i \cdot \boldsymbol{l}_j \right) \Psi(\mathbf{x}) \nonumber\\
&= \sum_{i=1}^A \left[ (l_i^x)^2 + (l_i^y)^2 + (l_i^z)^2 \right] \Psi(\mathbf{x}) + 2 \sum_{i<j} (l_i^x l_j^x + l_i^y l_j^y + l_i^z l_j^z) \Psi(\mathbf{x}).
\end{align}

The one-body part reads
\begin{align}
(l_i^x)^2 &= -y_i^2 (\nabla_i^z)^2 + 2y_i z_i \nabla_i^y \nabla_i^z - z_i^2 (\nabla_i^y)^2 + y_i \nabla_i^y + z_i \nabla_i^z, \nonumber\\
(l_i^y)^2 &= -z_i^2 (\nabla_i^x)^2 + 2z_i x_i \nabla_i^z \nabla_i^x - x_i^2 (\nabla_i^z)^2 + z_i \nabla_i^z + x_i \nabla_i^x, \nonumber\\
(l_i^z)^2 &= -x_i^2 (\nabla_i^y)^2 + 2x_i y_i \nabla_i^x \nabla_i^y - y_i^2 (\nabla_i^x)^2 + x_i \nabla_i^x + y_i \nabla_i^y.
\end{align}

The two-body cross terms are given by
\begin{align}
l_i^x l_j^x &= -y_i y_j \nabla_i^z \nabla_j^z + y_i z_j \nabla_i^z \nabla_j^y + z_i y_j \nabla_i^y \nabla_j^z - z_i z_j \nabla_i^y \nabla_j^y, \nonumber\\
l_i^y l_j^y &= -z_i z_j \nabla_i^x \nabla_j^x + z_i x_j \nabla_i^x \nabla_j^z + x_i z_j \nabla_i^z \nabla_j^x - x_i x_j \nabla_i^z \nabla_j^z, \nonumber\\
l_i^z l_j^z &= -x_i x_j \nabla_i^y \nabla_j^y + x_i y_j \nabla_i^y \nabla_j^x + y_i x_j \nabla_i^x \nabla_j^y - y_i y_j \nabla_i^x \nabla_j^x.
\end{align}


%